\documentclass[12pt,preprint]{aastex}

\newcommand{\source}{4U 1608--522}

\shorttitle{Spectral state variations of 4U 1608--522}
\shortauthors{Tarana A. et al.}

\begin{document}

\title{{\it INTEGRAL}{$^\star$} and {\it BeppoSAX} observations of the transient atoll source 4U 1608--522: from quiescent to hard spectral state}

\author{A. Tarana,\altaffilmark{1,2} A. Bazzano\altaffilmark{1} and
  P. Ubertini,\altaffilmark{1}}
 \altaffiltext{1}{Istituto di Astrofisica Spaziale e Fisica Cosmica-INAF,
via del Fosso del Cavaliere 100, I-00133 Roma}\email{antonella.tarana@iasf-roma.inaf.it}
\altaffiltext{2}{Dipartimento di Fisica, Universit\`a di Roma Tor Vergata, via della Ricerca Scientifica 1, I-00133 Roma}
\altaffiltext{$^\star$}{INTEGRAL is an ESA project with instruments and science data centre funded by ESA member states (especially the PI countries: Denmark, France, Germany, Italy, Switzerland, Spain), Czech Republic and Poland, and with the participation of Russia and the USA.}

\begin{abstract}
In this paper we report on the spectral evolution of 4U 1608--522 performed as part of the long Galactic Bulge monitoring with {\it INTEGRAL\/}. The data set include the April 2005 outburst. {\it BeppoSAX} archival data (two observations, in 1998 and 2000) have been also analysed and compared  with the  {\it INTEGRAL\/} ones. Three different spectral states have been identified from the hard Color--Intensity diagram derived from {\it INTEGRAL\/}: the canonical Hard and Soft as well as an Intermediate state.
The hard state spectrum is well described by a weak black body component plus a Comptonised plasma component with high electron temperature ($kT_{\rm e}\simeq 60$ keV) extending up to 200 keV without any additional cut--off.  
The soft spectra are characterized by a cold Comptonised plasma ($kT_{\rm e}$= 2--3 keV, and 7 keV for the intermediate state) and a strong disk black body component. A reflection component, indicating reflection  of the X-ray radiation from the accretion disc, is also present in the soft state revealed by  {\it BeppoSAX} in 1998. The 2000 {\it BeppoSAX} observation revealed the source in quiescent state modelled by a neutron star atmosphere (assuming a neutron star with radius 10 km and mass 1.4 M$_\odot$) with an effective temperature, $kT_{\rm eff}$ of 0.1 keV plus a power law component with $\Gamma \sim$ 3 detected for the first time for this source. This spectrum can also be modelled with a simple black body compatible with emission originating from a small fraction of the NS surface of radius of 0.4 km.
\end{abstract}

\section{Introduction}
A Low Mass X-ray Binary (LMXB) is a stellar system composed by a late type star and a compact object which accretes material from the companion star via Roche lobe overflow. The Neutron star (NS) LMXBs  are classified as Z or Atoll source based on the track of the colour-colour diagram \citep{hasinger} corresponding to different X-ray spectral and variability properties. The Atoll sources move from the Island branch that corresponds to the Hard spectral state to the Banana branch corresponding to the Soft state.

Many LMXBs are Soft X-ray Transient sources (SXRTs) showing spectral state changes driven possibly by accretion instability phenomena. In fact, they are often in a quiescent state with a low luminosity of 10$^{33}$ erg s$^{-1}$ and abruptly show X-ray outbursts that consist of fast flux increase reaching a $L_X \sim 10^{37-38}$ erg s$^{-1}$ followed by a slower nearly exponential decay (lasting weeks or months) \citep{campanareview}. During the outburst the sources reveal spectral states transitions from the Hard (onset the outburst) to the Soft (at the peak of the outburst) passing through Intermediate spectral states. The mass accretion rate is the main parameter responsible for the spectral changes (hardening and softening).  NS in quiescent state on average are more luminous than black holes (BH) in quiescence. This observational property may be used to distinguish the NS or BH nature of the compact object (Rutledge et al. 2000, Campana et al. 2001). Unfortunately so far, few sources have been detected during the quiescent state namely Aql~X-1, Cen~X-4, EXO~0748--676, 4U~2129+47 \citep{campanareview}, XTE~J2123-058 \citep{tomsick04} and 4U~1608-522 itself \citep{asaiquie}.

The Atoll 4U~1608--522 is a LMXB with a NS as compact object as derived by the presence of Type I X-ray bursts and superbursts (Nakamura et al. 1989, Remillard et al. 2006). It is one of the SXRTs showing long periods of quiescence (such us EXO~0748--676 and 4U~2129+47) spaced out by periodic outbursts (Lochner et al. 1994, Simon et al. 2004) like Aql~X-1 \citep{tanaka}.

In the past decade, the {\it BeppoSAX\/} and {\it RXTE\/} broad band capability provided a step forward in the study of the NS LMXBs establishing their high energy behaviour well above the standard X-ray band ($\le$ 20 keV). More recently, reflection components and high energy tails have been reported, e.g. 4U~1820--30, 4U~1705--44  (Tarana et. al. 2007, Fiocchi et al. 2006, Piraino et al. 2007) and 4U~1608--522 itself \cite{gie}. For \source\/ a radio flux density upper limit of $<$0.19 mJy at 8.5 GHz is available \cite{migliari06}.

We report here on recent data from {\it INTEGRAL\/} to characterise the high energy emission and spectral variation. For this purpose we performed the scientific analysis of the data in the 4-200 keV energy range with JEM-X \citep{lund} and IBIS \citep{uber}. Two non contemporaneous observations with {\it BeppoSAX\/} \citep{boella} have been also analysed.

We combined three methods of analysis: {\it Temporal}, using the light curves in different energy bands; {\it Photometric}, with the hardness--instensity Diagram; and {\it Spectral}, by spectral modelling.

\section{Observations and data analysis}


The {\it INTEGRAL\/} observations span a non continuous period starting from 2004 February 18 to 2005 September 8, during the revolutions 164--354. We analysed all the public data of JEM-X (within $7\degr\times 7\degr$ of the FOV) and IBIS (within $9\degr\times 9\degr$ FOV) which resulted in 192 and 420 pointings respectively (Science Window, ScW), each lasting about 2000 seconds. Data were extracted with the Off-Line Scientific Analysis (OSA) \citep{gold} v.\ 5.1 software released by the {\it INTEGRAL\/} Science Data Centre (Courvoisier et al.\ 2003).
For the spectral analysis, performed with the XSPEC
package v.\ 11.3, 2\% systematic errors to both JEM-X and IBIS data sets have been
added\footnote{http://isdc.unige.ch/?Support+documents}. Fluxes have been
normalized to IBIS/ISGRI value.

The two public observations of {\it BeppoSAX\/} (available at http://asdc.asi.it)  were performed on 1998 February 28 and 2000 August 1. The LECS (0.1-10 keV), MECS (1-10 keV) and PDS (15-200 keV) spectra have been generated within a radius of  4$^{\prime}$ using  the publicly distributed responce matrices; a 1 \% systematic error has been added to the model.

\section{Light curves and hardness--intensity Diagrams}

\begin{figure*}[ht]
\centering
\includegraphics[height=10cm,angle=90]{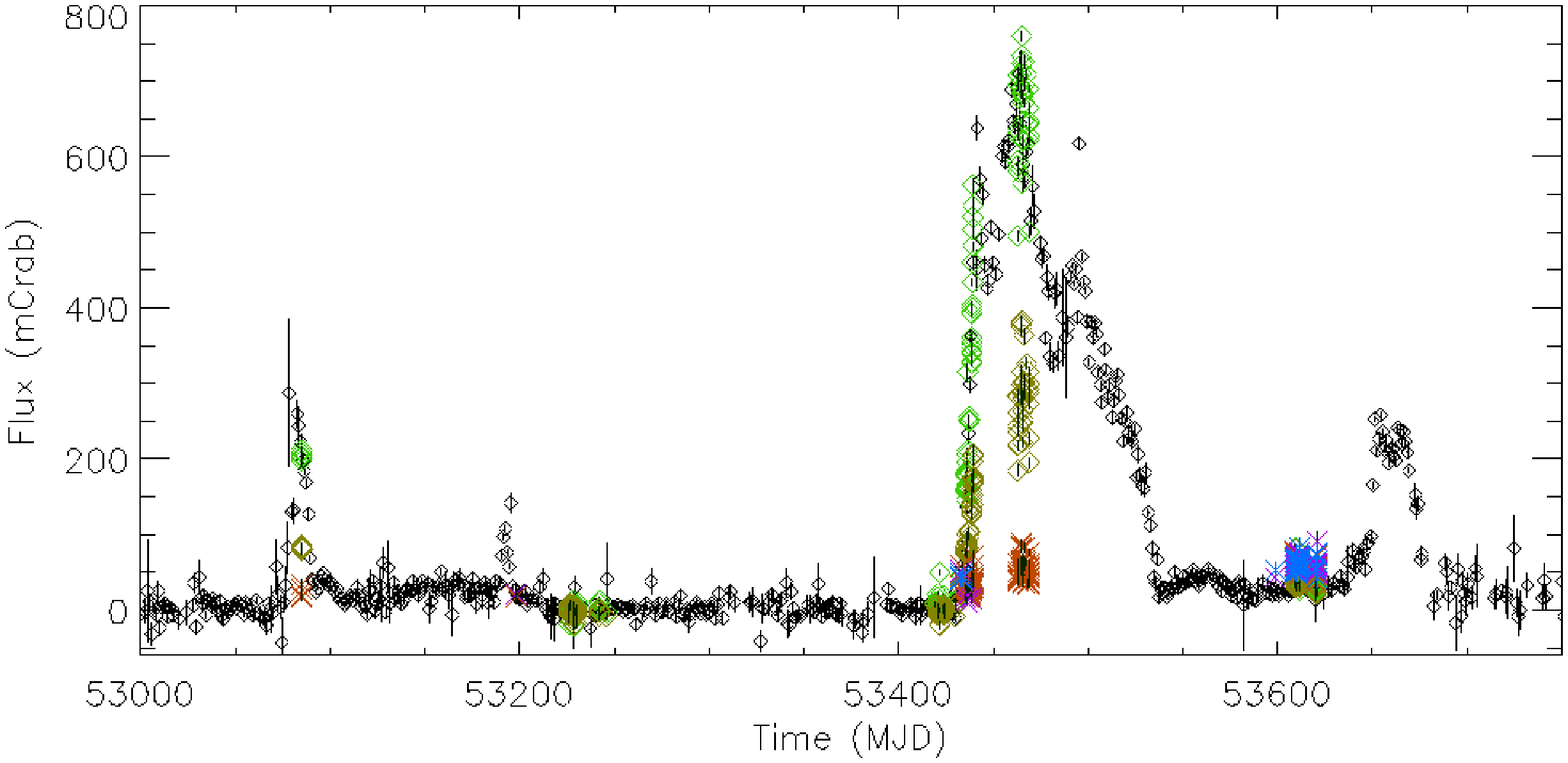}
\caption{The 2004--2005 light curves of \source\  with {\it RXTE}/ASM (black diamonds), {\it INTEGRAL}/JEM-X (light green diamonds in the 4-10 keV and in the dark green diamonds in the 10-20 keV band) and IBIS (red crosses in the 20-30 keV, purple in the 30-60 keV and blue in the 60-120 keV). Each point correspond to the count rate in  mCrab unit, for a time bin of 1 ScWs for {\it INTEGRAL\/} and 1-day everage for ASM. \label{plotone}}
\end{figure*}

\begin{figure*}[ht]
\centering
\includegraphics[height=10cm,angle=90]{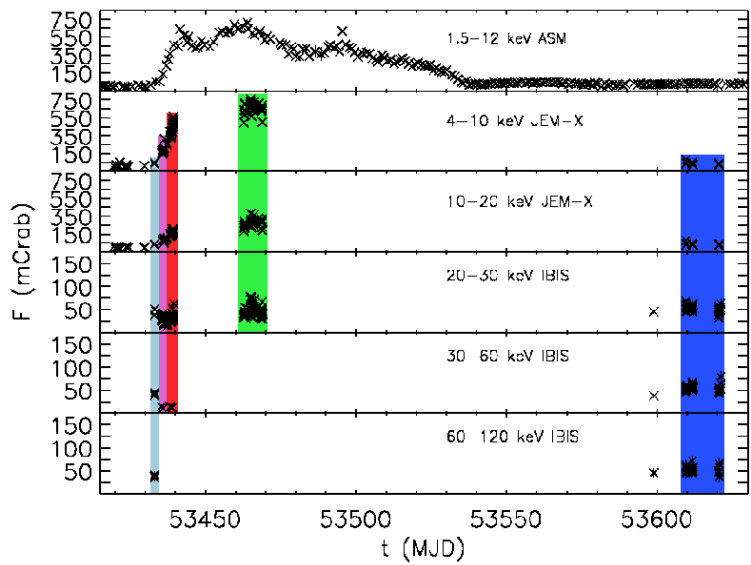}
\caption{The zoom during the outburst of February--April 2005. The colored lines correspond to different spectral states. Note that the hardening appear just at the beginning and the end of the outburst (see the blue crosses). These data are reported with the same colors in the hardness-intensity diagrams of Fig. \ref{plottre}.\label{plottwo}}
\end{figure*}

\begin{figure*}[ht]
\centering
\includegraphics[height=8cm,angle=90]{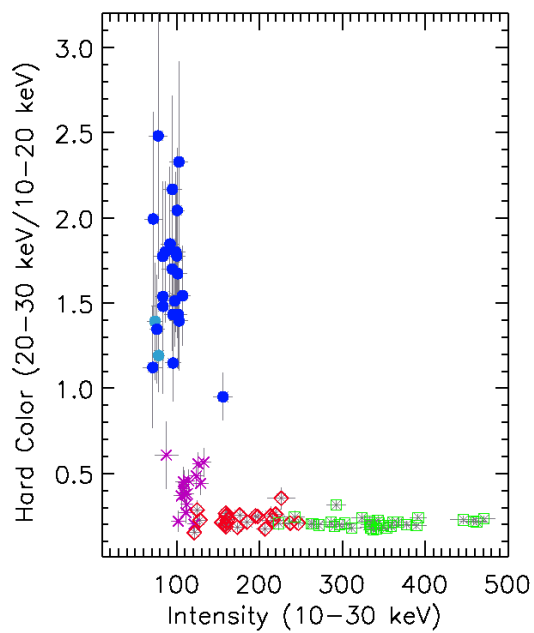}
\caption{Hard color--intensity diagram with IBIS/JEM-X data. 
Each point corresponds to a 1 ScW.
\label{plottre}}
\end{figure*}

In Fig.\ \ref{plotone} we report the public 1.5-12 keV {\it RXTE}/ASM\footnote{http://xte.mit.edu/ASM$\_$lc.html.} light curves of \source\  in the period  2004 February--2005 September and overplot the light curves for JEM-X (4--10, 10--20 keV bands) and for IBIS (20--30, 30--60, 60--120 keV bands). Each {\it INTEGRAL\/} point corresponds to a  single ScW. Fig.\ \ref{plottwo} is a zoom  of the light curve during the outburst of the source started on February 2005 (MJD 53400). Different colors in Fig.\ \ref{plottwo}  correspond to different spectral properties. Colors in the hardness--intensity diagrams of Fig.\ \ref{plottre} have been used accordingly. This figure shows the hard color--intensity diagram from  JEM-X/IBIS data. Hard color here are defined as the ratio of the 10--20 to 20--30 keV band fluxes, while the Intensity are the sum of the 10--20 keV plus 20--30 keV band fluxes (all fluxes are in mCrab units). 
The dark and light blue data correspond to a hard spectral state (island state)  while the purple red and green data correspond to soft states (banana states). The hard spectral states appear at the beginning and the end of the outburst (see Fig. \ref{plottwo}) as typical for transient sources.

\begin{figure*}[ht]
\centering
\includegraphics[height=10cm,angle=90]{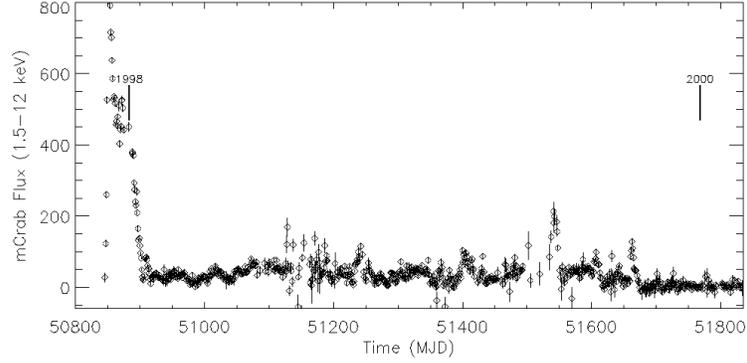}
\caption{ASM 1.5-12 keV light curve, in which are indicated with vertical lines the BeppoSAX observations. \label{plot_sax}}
\end{figure*}

In Fig. \ref{plot_sax} we show the ASM light curve with the two {\it BeppoSAX\/} observations marked with vertical lines. In the 1998 observation the source was in a soft state, while in 2000 it was in quiescent phase, as discussed below.

\section{Spectral analysis}
\label{spectral}
Spectra for {\it INTEGRAL\/} have been derived with pointings shown with the same color in  the hardness--intensity diagram (see Fig \ref{plottre}). In Table \ref{tab_dataSET} we reported the log of the {\it INTEGRAL\/} and {\it BeppoSAX\/}  data sets and spectral states. Data S1 and S5 correspond to Hard state; S3 and S4 correspond to Soft state; and S2 to the soft/intermediate state.

\subsection{Soft and Intermediate states}
S2, S3 and S4 spectra have been first modelled with a Comptonised component, \texttt{CompTT} in XSPEC \citep{sun}. For the S2 and S3 spectra and, to minor extent, for S4 it was necessary to add a soft component  because of a soft excess below 6 keV and the poor values of the reduced $\chi^2$ ($\chi_{r}^2$ is 1.4 and 1.2 respectively). A multicolor disk black body component, \texttt{diskbb} in XSPEC \citep{mitsuda}, has been used though a black body component (\texttt{bb}) gives similar results. In Table \ref{tabspe_soft}, the parameters of the \texttt{Comptt+diskbb} model  have been listed for these spectral states.

The Comptonization model has three main parameters: $kT_{\rm e}$, the temperature of the electrons of the corona i.e. a hot region near the disk; the optical depth of this region, $\tau$; $kT_{\rm 0}$, the temperature of the input seed photons that are upscattered to higher energies by the hot electrons of the corona. The temperature of the disk black body depends on the radius as $T\propto R^{-\frac{3}{4}}$, and the fit parameter of the diskbb model, $kT_{\rm in}$, is the temperature at the inner disk radius, $R_{\rm in}$, corresponding  to the inner last stable orbit of the material of the disk.

The fit results of \texttt{CompTT+diskbb} model of Table \ref{tabspe_soft}, show that the temperature of the Comptonised region ($kT_{\rm e}$) decreases from the left banana to the right banana (S2 to S4) and becomes optically thick ($\tau$ increases), whereas the inner disk black body temperature ($kT_{\rm in}$) indicates a small increase.
Moreover, for the spectra S2 and S3, the temperature of the input photons required for the Comptonization, $kT_{\rm 0}$, has a different value from the  inner disk temperature of the photons, $kT_{\rm in}$. This indicates that the seed input photons should come from a region with a higher temperature respect to the inner disk region, such as boundary layer or from the neutron star surface. On the contrary, for S4 the temperature of the inner disk is compatible (within the errors) with the $kT_{\rm 0}$  value. In the fitting procedure, when trying to fix the parameters $kT_{\rm 0}$ and $kT_{\rm in}$ at the same value for all spectral data, only the spectrum S4 gives an acceptable $\chi_\nu^2$ value. On the other hand, we are aware  that this parameter is well out of the instrument energy coverage.


A better estimate of the soft state parameter have been obtained from a  {\it BeppoSAX\/} observation in 1998 that covers energies well below 4 keV (below JEM-X operative range). Results of the fit with the addition of an absorption model with a column density $N_{\rm H}$ of $1.2\times 10^{22}$ cm$^{-2}$ (compatible with previous found by Penninx et al. 1989) is reported in Table \ref{tabspe_soft}. Also in this occasion the source reveal a temperature of the input photons higher than the temperature of the inner disk photons temperature. We further discuss this in the Discussions and Conclusions section.

\begin{figure*}[ht]
\centering
\includegraphics[height=7.6cm,angle=-90]{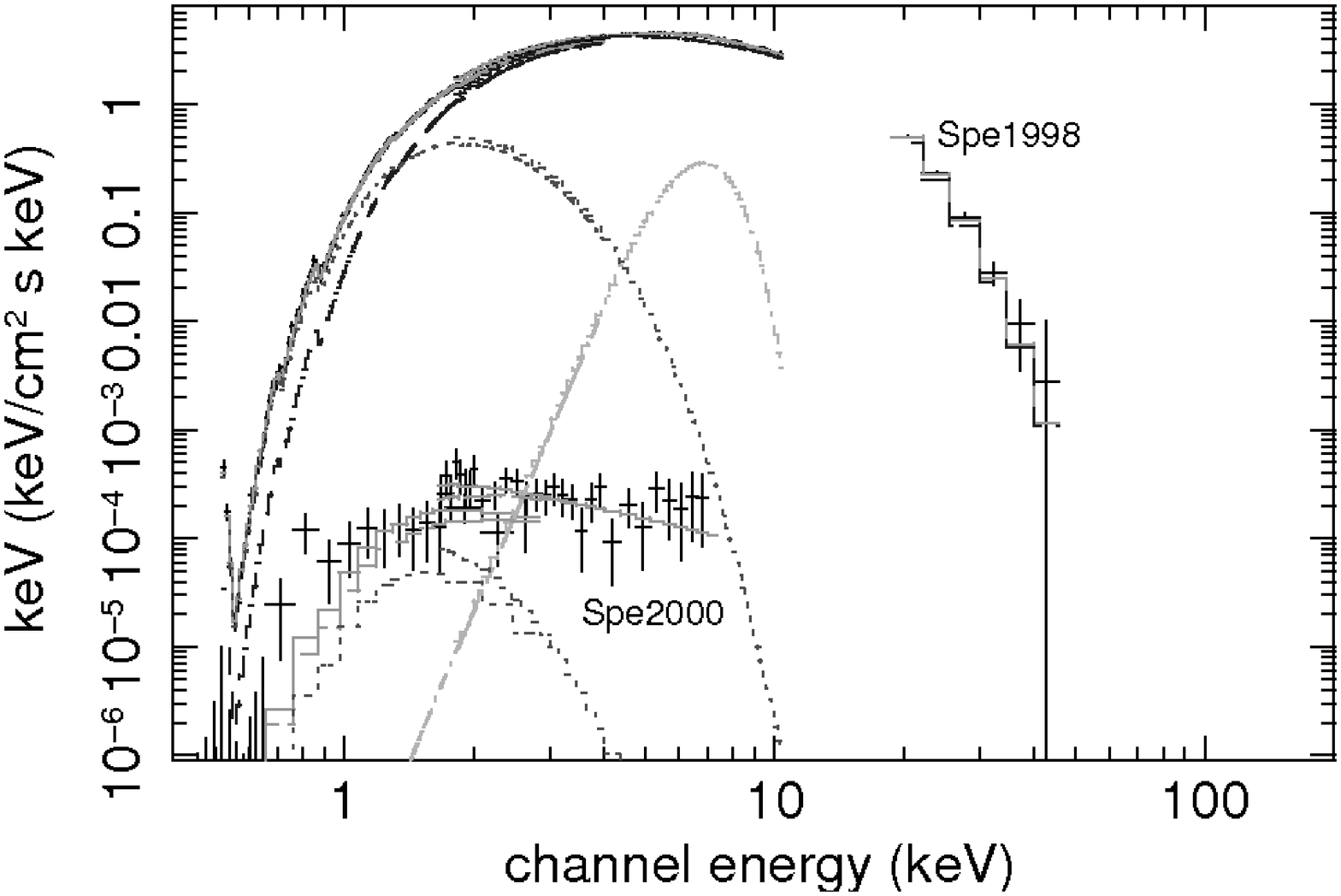}
\includegraphics[height=7.6cm,angle=-90]{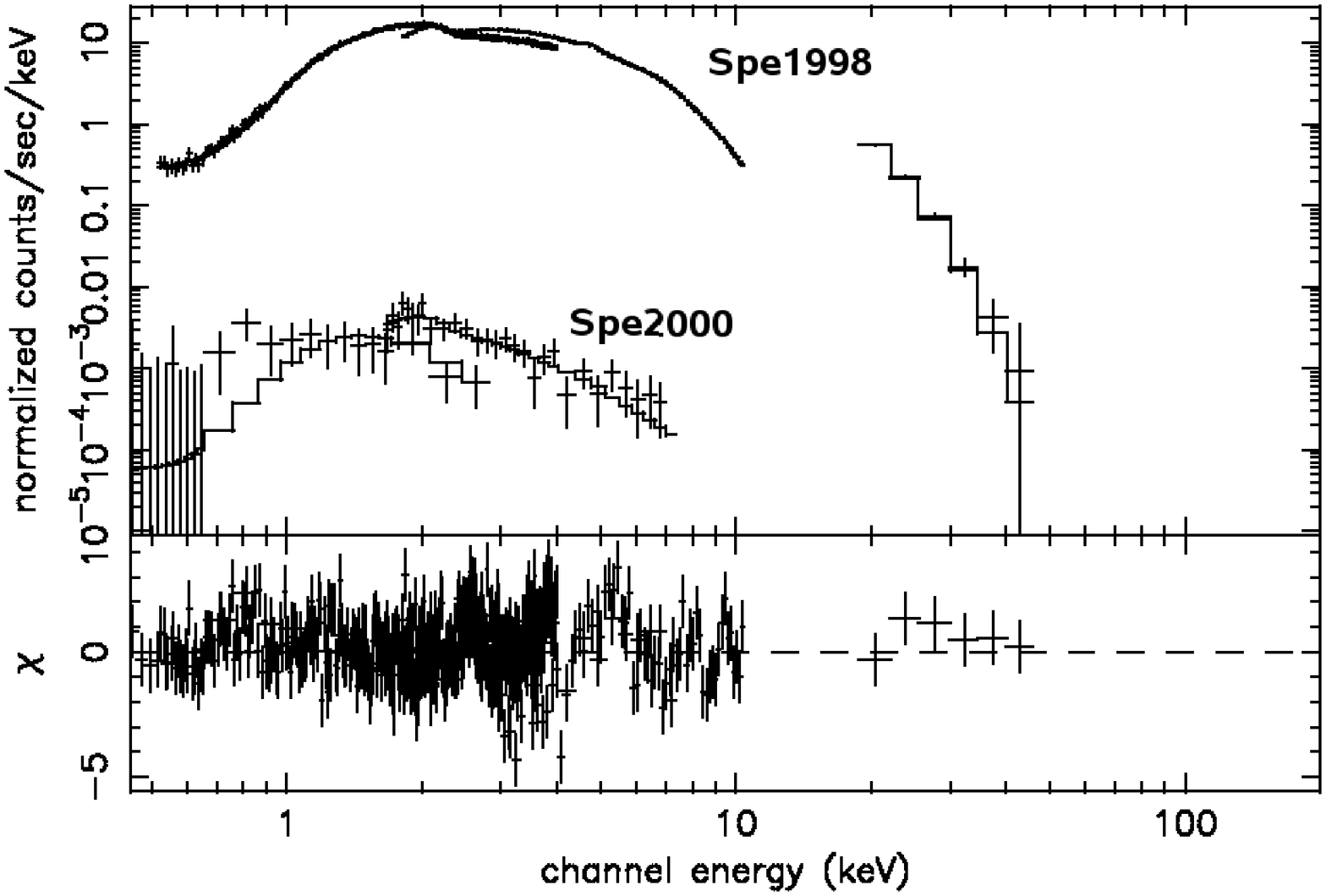}
\caption{\emph{BeppoSAX} soft and quiescent spectra of 4U 1608-522. Spe1998: soft spectrum  with CompTT+diskbb+reflect+gaussian model. Spe2000: quiescent spectrum modelled by NSA plus power law models.\label{plotfive}}
\end{figure*}

A reflection component \citep{ward} has been tentatively added to all the observed soft states and it is required only for the {\it BeppoSAX\/} soft state data. In fact, for S3 and S4 spectrum the F-test does not improve, whereas for S2 spectrum a better fit was obtained but with a low chance probability (10\%). 
The reflection component contributes to the fit with a sort of bump in the 10-20 keV band and as consequence we get a better fit of the high energy data ($>$ 30 keV). Moreover the best fit parameters value doesn't really change from the previous fit. Parameters are shown in Table \ref{tabspe_soft} and  the energy and counts spectra are showed in  Fig. \ref{plottre}. For the BeppoSAX data we need to add also a gaussian model to fit an excess below 10 keV, as it is showed in Fig. \ref{plotfive}, where also the quiescent spectrum is reported. A Fe reflection line at energy of 6.4 keV with a large FWHM is required to obtain the best fit model.

Assuming the source distance at 3.6 kpc \citep{nakamura}, the value of the bolometric luminosity ranges from 6 to 1.5$\times 10^{37}$ erg s$^{-1}$ from the soft to the intermediate state respectively.

\subsection{Hard states}
Hard states have been detected at the beginning and at the end of the 2005 outburst as derived from light curves and hardness--intensity diagram shown in Figures \ref{plotone}, \ref{plottwo} and \ref{plottre}. 
The S1 and S5 data sets was modelled by the same model with the same spectral parmeters so that have been added in a single spectrum to increase the signal to noise ratio.

\begin{figure*}[ht]
\centering
\includegraphics[height=8cm,angle=-90]{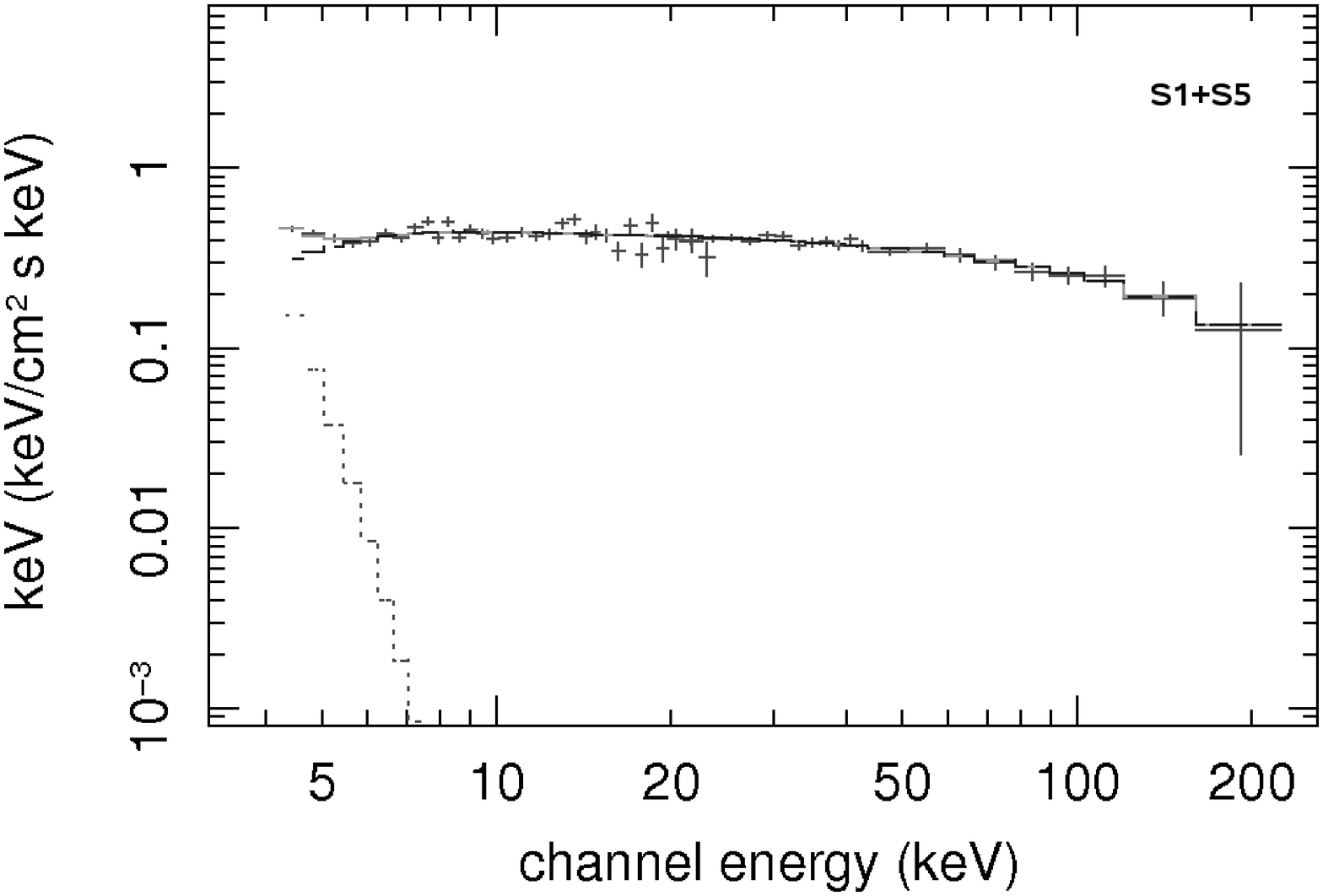}  
\includegraphics[height=8cm,angle=-90]{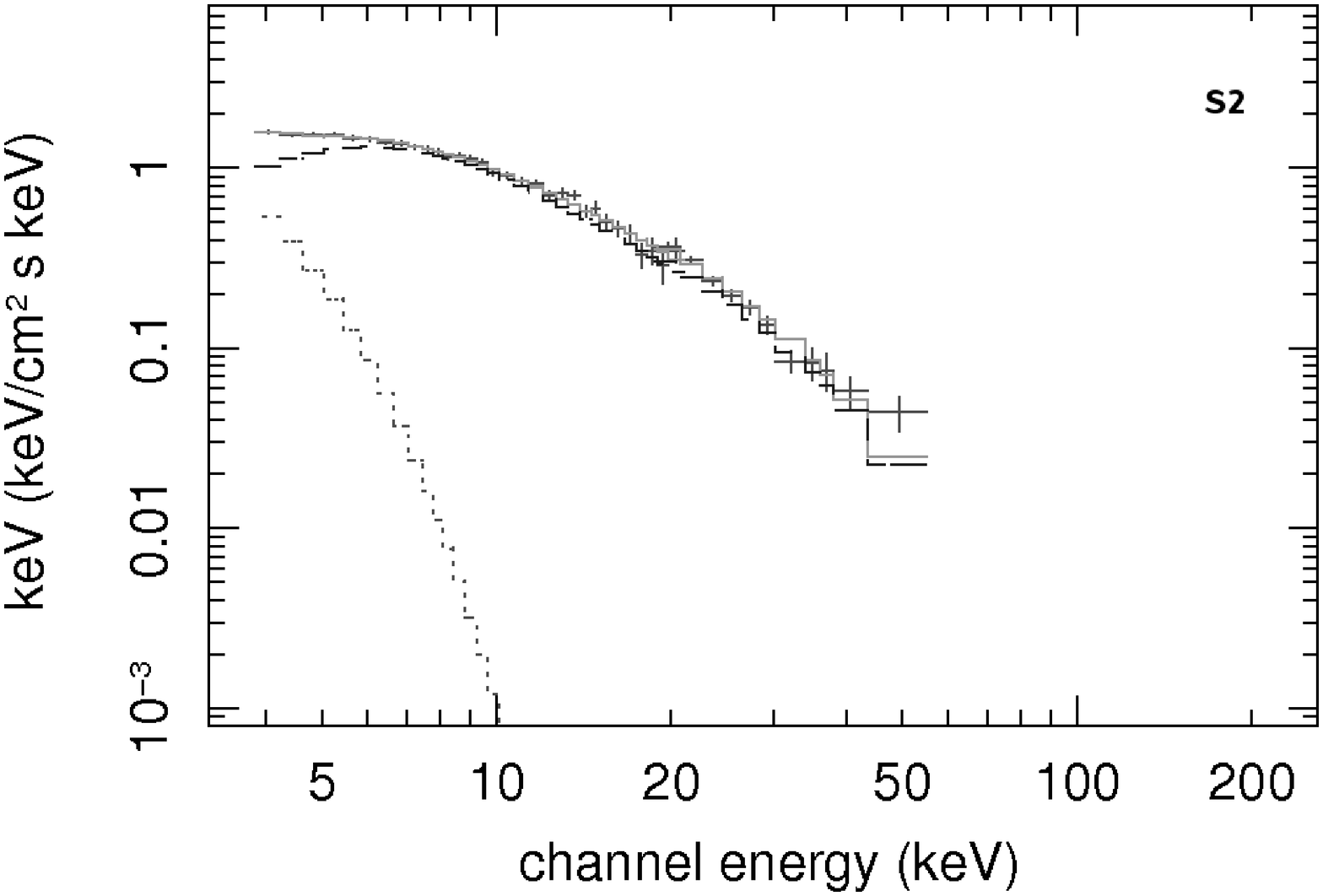}  
\includegraphics[height=7.8cm,angle=-90]{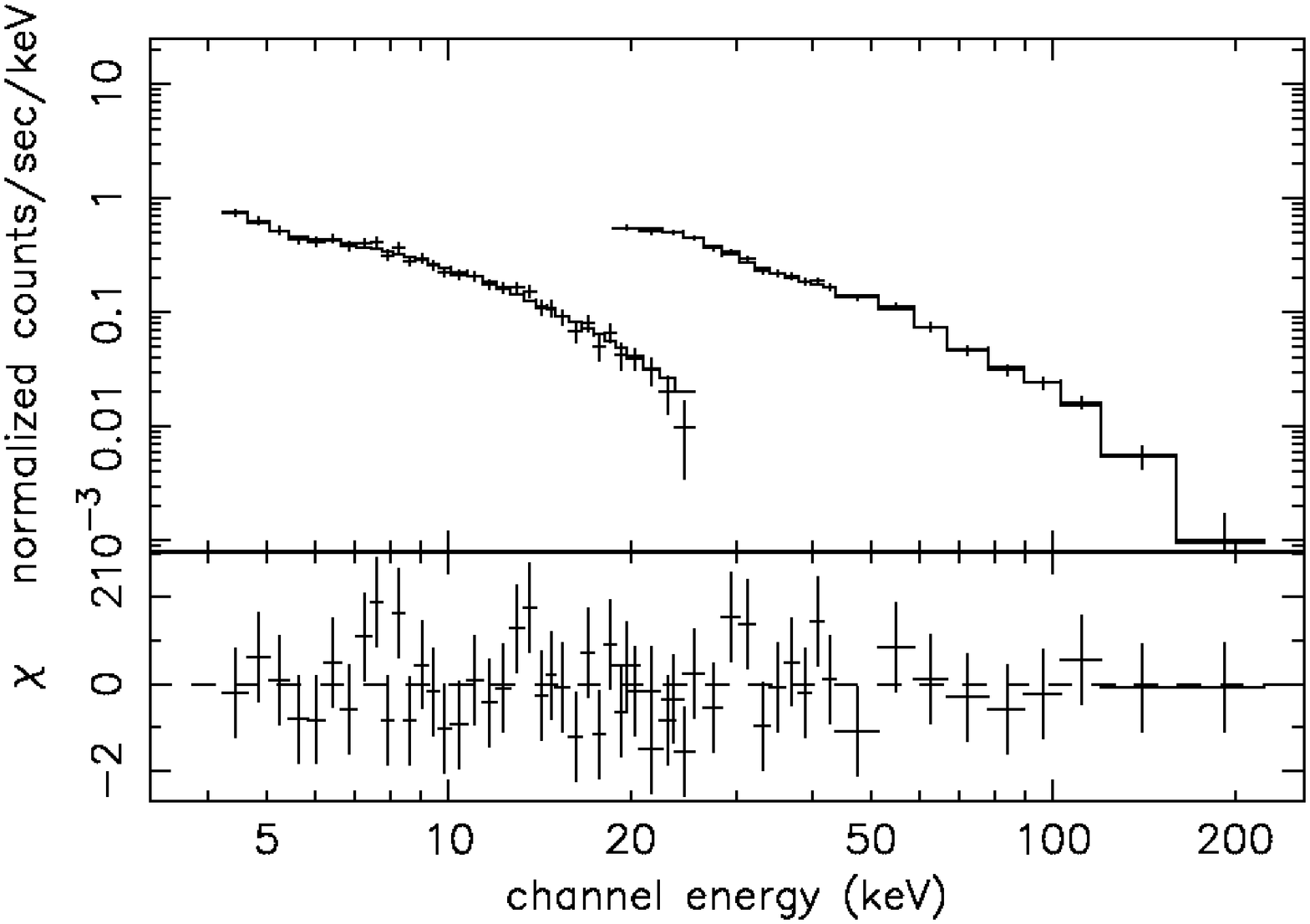} 
\includegraphics[height=7.8cm,angle=-90]{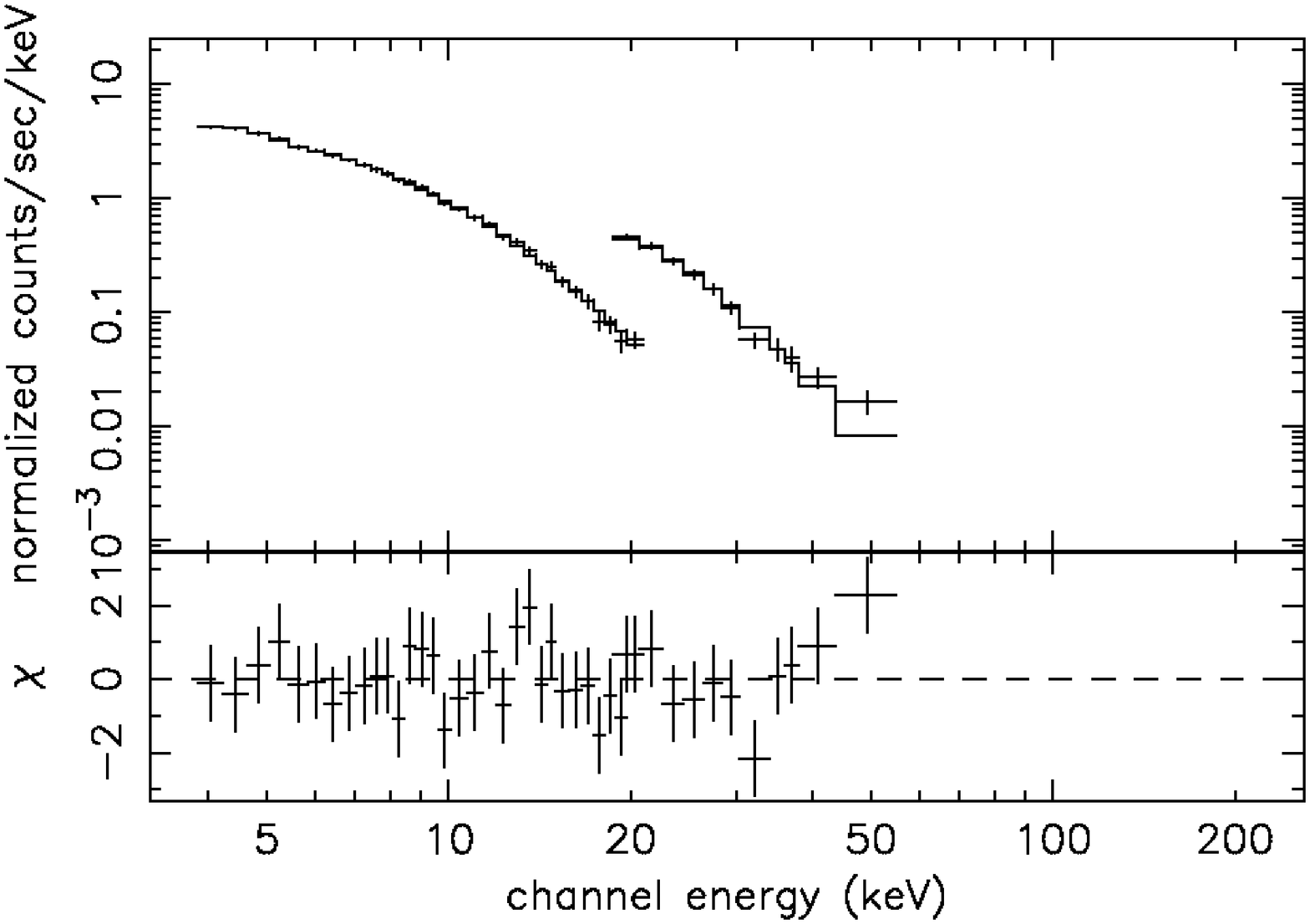} 
\includegraphics[height=8cm,angle=-90]{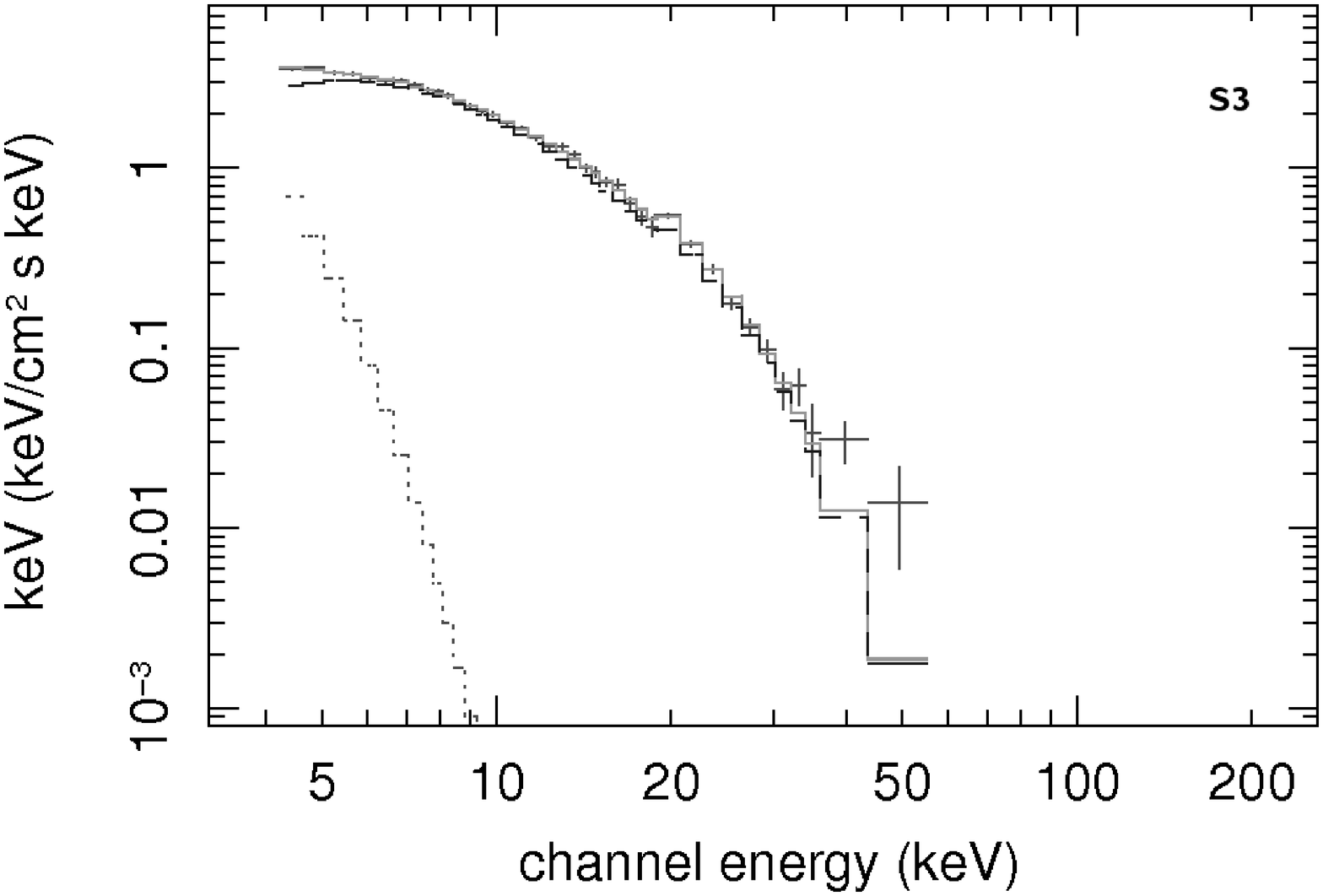} 
\includegraphics[height=8cm,angle=-90]{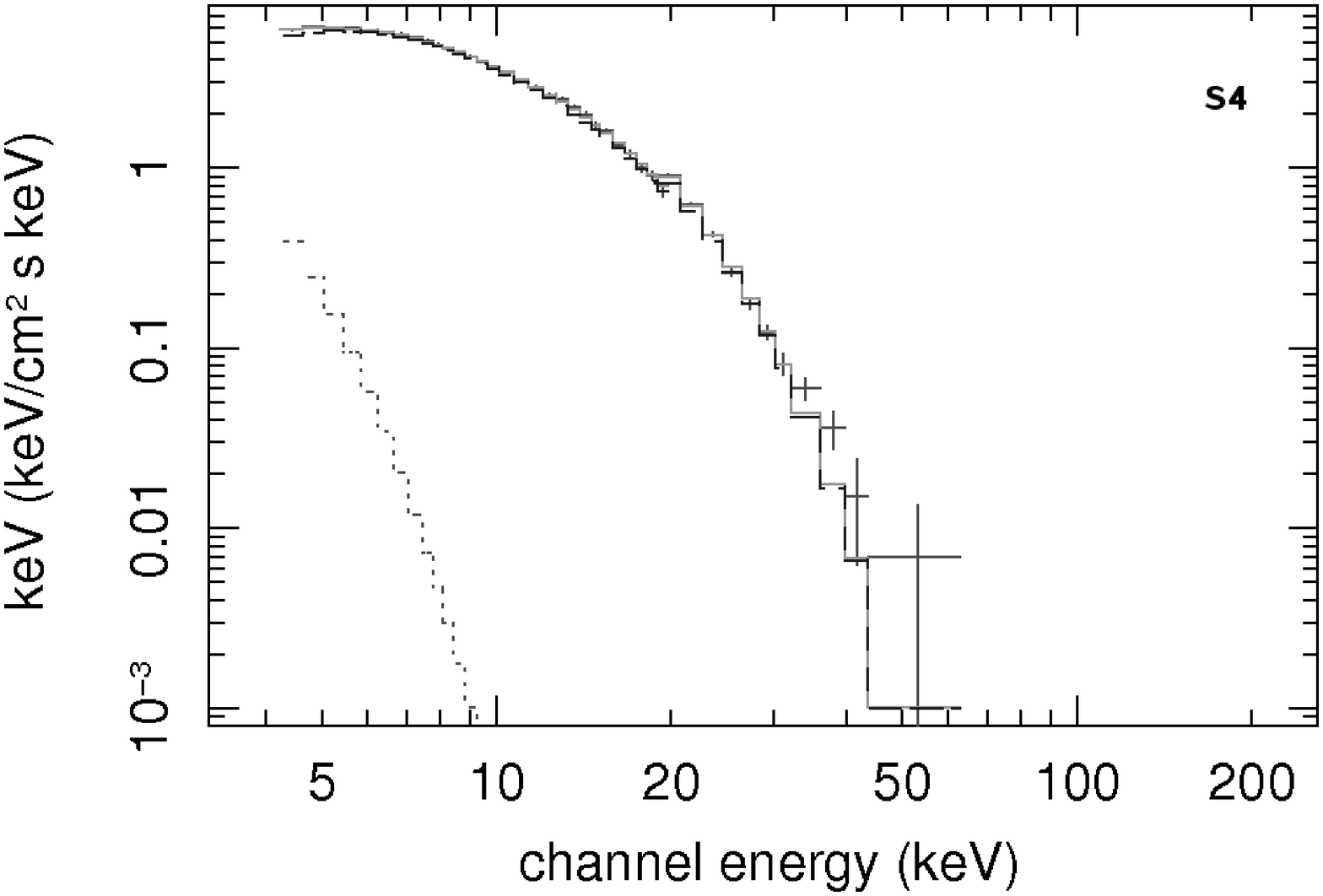} 
\includegraphics[height=7.8cm,angle=-90]{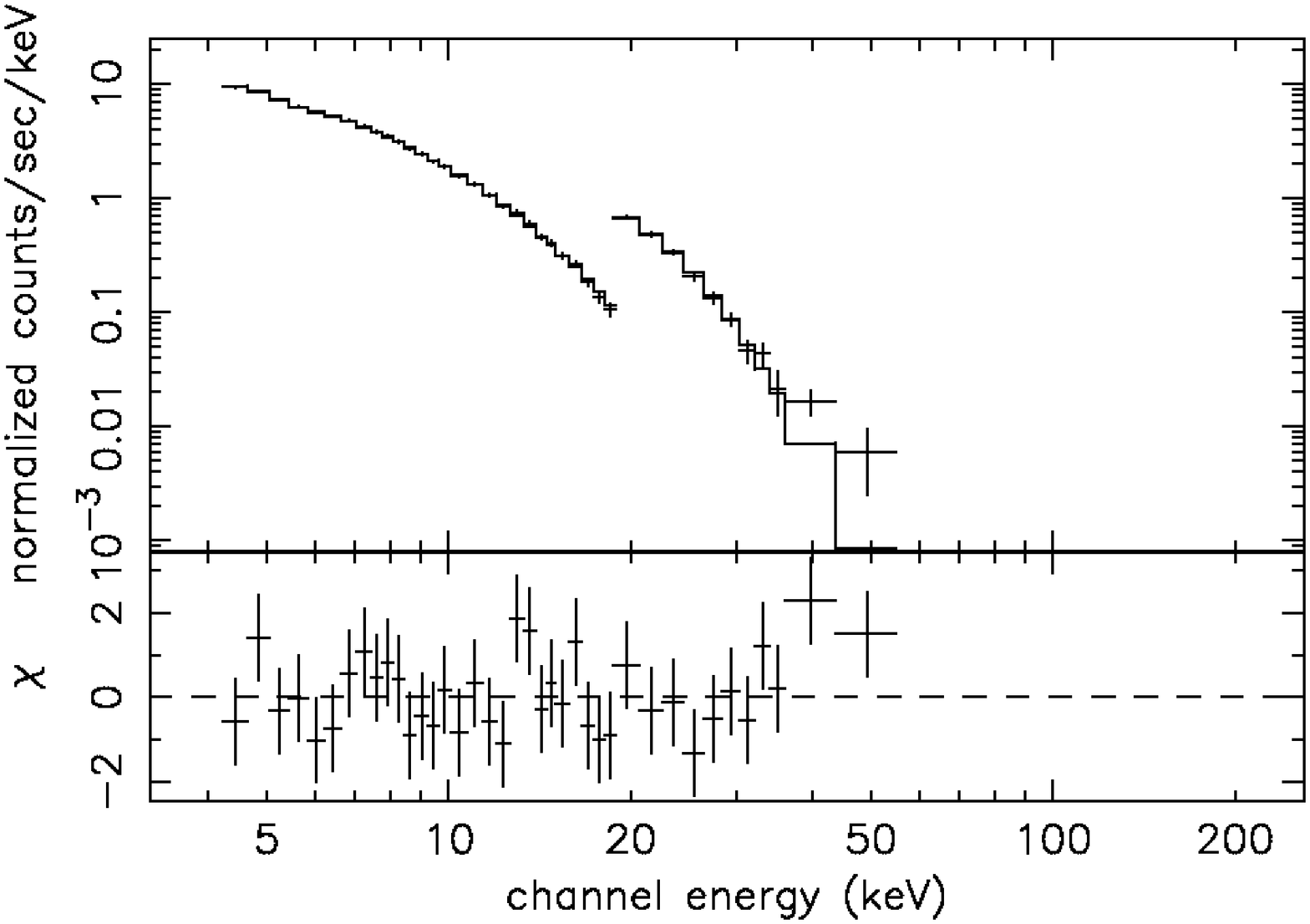}
\includegraphics[height=7.8cm,angle=-90]{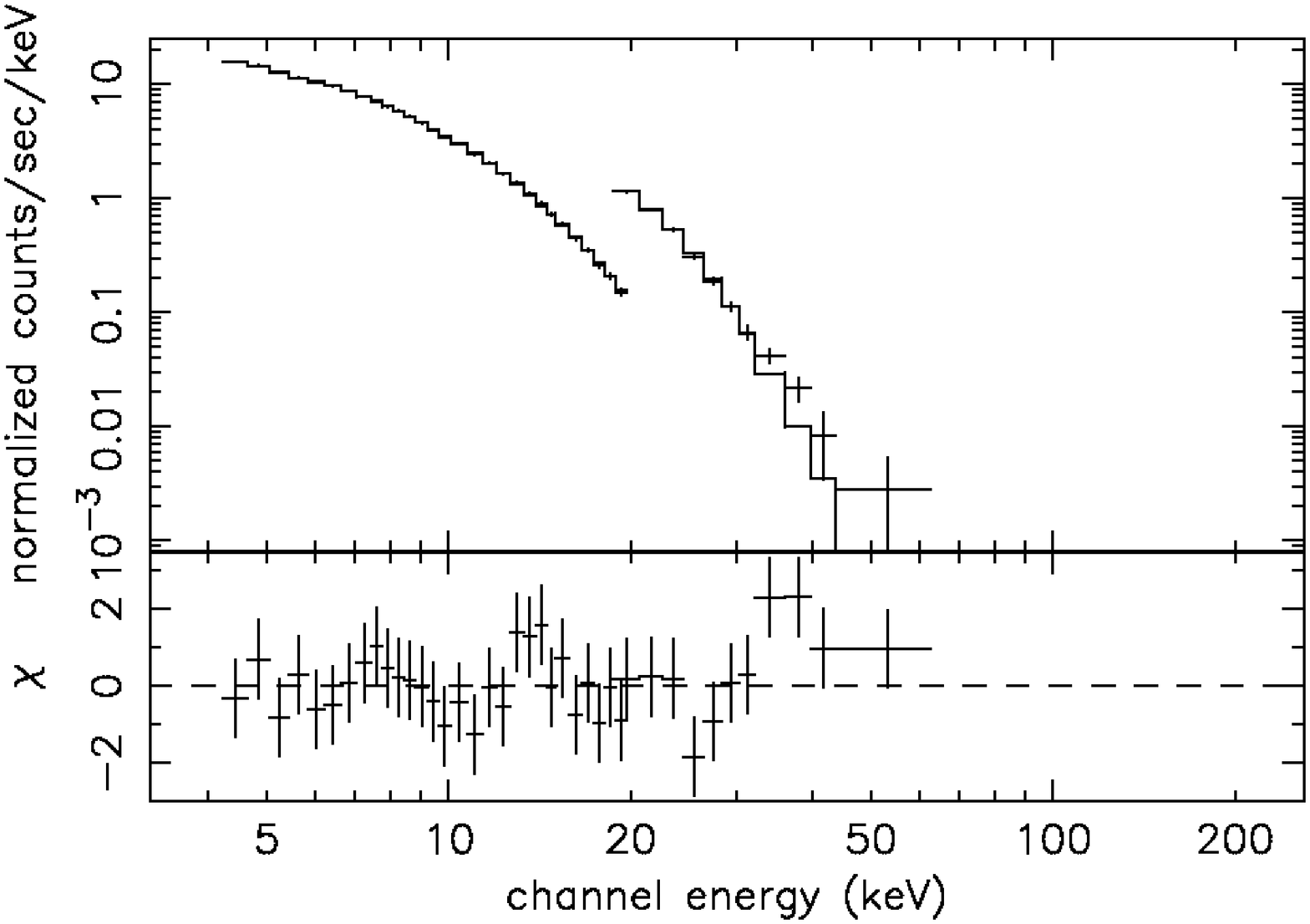}
\caption{{\it INTEGRAL} spectra of 4U~1608--522 during the 2005 outburst. From the top left to the bottom right: Hard (S1+S5 data sets) and Soft spectra (S2, S3, S4 data sets) from IBIS and JEM-X, with the CompTT+diskbb+reflect models.\label{plotfour}}
\end{figure*}

Firstly a Comptonization model has been used to fit the data and parameters are: $kT_{\rm e}$=57.4$^{+54.3}_{-21.3}$, optical depth, $\tau$=0.49$^{+0.41}_{-0.31}$, and input photon temperature, $kT_{\rm 0}$=0.5 keV, with a $\chi_r^2$($\nu$)= 0.91(50). A \texttt{diskbb} component has been added to take into account  the higher residuals below 6 keV. Table \ref{tabspe_hard} reports the obtained values of this fit, and Fig. \ref{plotfour} shows the spectrum extending up to 200 keV. 
Moreover, an acceptable fit has been obtained also freezing the temperature of the seed photons to the black body temperature. In this case a soft temperature is 0.5 keV, and the corona temperature and optical depth are similar to the previous model.

It is evident in all cases that the value of the temperature of the electrons of the corona is not well costrained, indicating that a real cut--off in this spectrum is not detected.
We also checked the SPI data in the same period and there is agreement with our ones [E. Jourdain, private comunication].

Finally we try to fit this spectrum with the \texttt{CompPS} model \citep{compps}, which include the reflection parameter, with no improvement as showed in Table \ref{tabspe_hard}.

The unabsorbed bolometric luminosity of the hard spectrum correspond to 5$\times 10^{37}$ erg~s$^{-1}$.

\subsection{Quiescent state}
During the 2000 observation of {\it BeppoSAX\/} the source was in a quiescent state as derived with LECS and MECS only and the  bolometric luminosity correspond to $\sim$1.4 $\times 10^{33}$ erg~s$^{-1}$. 

This quiescent state was fitted with different models: a simple power law, a black body, a black body plus a power law, a neutron star atmosphere model (\texttt{NSA} of XSPEC) \citep{nsa}, and finally a NSA plus power law. For all the fits we assumed an absorption, $N_{\rm H}$, fixed to 1.2$\times 10^{22}$ cm$^{-2}$.

The single power law component gives a photon index of  3.3 with a $\chi_r^2$($\nu$) value of 0.82(65). The black body model gives a temperatures 0.5 keV, with $\chi_r^2$($\nu$)equal to 1.18(65). From the value of the  blackbody normalization (assuming a distance of 3.6 kpc) we derive the black body emission region as $\sim$0.4 km that is incompatible with the typical neutron star radius of 10 km. This could be explained assuming that the X-ray emission originates from a smaller emission region such as the polar caps of the neutron star.
A better fit is obtained by adding a power law component [$\chi_r^2$($\nu$)=0.74(63)] but this imply a black body temperature of 0.07 keV and a black body region emission larger than the neutron star radius ($\sim$ 100 km).

We also try to fit the quiescence spectra with  a pure H-atmosphere spectrum. The Neutron Star H Atmosphere model (NSA) uses the mass and radius of the neutron star and the unredshifted effective temperature of the surface of the neutron star ($kT_{\rm eff}$) as parameters. The model normalization is equal to $1/D^2$ where D is the distance of the source. Note that the black body model temperature (the temperature to infinite distance) is different from the $kT_{\rm eff}$ by the redshift factor that is 0.76 for the standard value of mass and radius of a NS (1.4M$_\odot$ and 10 km). We froze the mass and the radius at the standard value of a NS and the normalization to the value of 7.716$\times 10^{-8}$ (for D equal to 3.6 kpc) and let the effective temperature as free parameter. The NS radius has been frozen at 10 km to overcome the underestimate of the emission region resulting from black body model, as has been done for previous  NS quiescent state data \citep{rut99}, \citep{tomsick04}. The obtained effective temperature of the neutron star is 0.17 keV, but the value of the $\chi_r^2$($\nu$) is still too high [1.7(66)].


The best fit model has been obtained adding a power law to the NSA component: in this case the effective temperature corresponds to 0.12 keV, and the photon index of the power law to 3 [$\chi_r^2$($\nu$) is equal to 0.83(64)].

The 0.5-10 keV unabsorbed luminosity is 4 $\times 10^{33}$ erg s$^{-1}$ which is higher than previous {\it ASCA\/} measurements in quiescence 1.9 $\times 10^{33}$ erg s$^{-1}$ \citep{asaiquie} and 7.3$-$8.3 $\times 10^{32}$ erg s$^{-1}$ \citep{rut99}. These authors also modelled the quiescent spectrum of this source with the same models (black body, single power law or NSA models) though a power law component was never required to the fits. 

\section{Discussion and conclusions}

We reported on the study of the spectral behaviour, as derived with {\it INTEGRAL} and {\it BeppoSAX\/} satellites, of the SXRT source 4U 1608--522 in different spectral states characterised by luminosity ranging from 4 $\times 10^{33}$ erg s$^{-1}$ to 6 $\times 10^{37}$ erg s$^{-1}$ from the quiescent up to the soft state. The soft and hard spectra have been modelled with a soft component, described with a disk black body emission or black body emission, plus a Comptonization component which is described by a Comptonised corona. The {\it BeppoSAX\/} soft state spectra show also a reflection emission from the disk. 

During the hardening (from the banana to the island branch of the color-intensity diagram) the inner disk black body temperature decreases from 0.7 to 0.4 keV 
while the Comptonised corona temperature increases and the optical depth decreases, becoming an optically thin and very hot corona in the hard state ($kT_{\rm e}$ changes from 2 to 60 keV).

In spite of the difficulty to precisely determine the soft parameters  (because of the limited instrumentation band-width often sensitive at E$>$ 4 keV) and the need to always freeze the seed photons temperature in our fits, we can still draw some remarks and explanations on the source.

From the diskbb normalization model we have estimated the inner disk radius\footnote{estimated with formula given in Gienrli{\'n}ski \& Done 2002.}. For the soft spectra a radius ranging from 20 to 60 km has been obtained, indicating that the disk extend near the NS. On the contrary, for the hard spectrum the inner radius extends up to 120 km, as expected to allow the formation of hot plasma corona. This imply that change in the spectral state corresponds to a different disk-corona geometry  of the system, as already observed in NS and also BH transient systems \citep{zdz2004}. 

Moreover, we observed a seed input photons temperature higher ($T_{\rm 0}$ ranging from 0.6 to 1.2 keV) than the inner disk temperature. In view of the fact that the $kT_{\rm in}$ is representative of the latest/inner part of the accretion disk,  the seed photons temperature  must be provided from a hotter inner region that could be either close the neutron star i.e. an optically thick  boundary layer heated by friction effect (as suggests possibly for the source GRS 1724-30, Barret et al. 2000) or from the neutron star surface itself. 

The boundary layer model predicts a temperature ranging from 0.5 to 2.0 keV (Lin et al. 2007 and references therein) whereas the predicted temperature for the "spreading boundary layer model" of a NSLMXB (with NS radius of 15 km and mass of 1.4 M$_{\odot}$) is of 2.4 keV (Lin et al. 2007, Revnivtsev et al. 2006, Suleimanov \& Poutanen 2006) well above the temperature of 0.6-1.2 keV observed for 4U~1608--522.  Another possibility is that the seed photons could be provided by the stellar surface of the NS. In this case the temperature is higher than the NS temperature estimated during quiescence due to the NS cooling during the quiescent state at low accretion rate \cite{Done_review}. 

In the hard state the energy spectrum doesn't show any high energy cut--off. This behaviour is usually attributed either to non-thermal emission processes probably related to jet formation, as for 4U~0614+091 \cite{migliari06b}, or to hybrid thermal-non thermal emission, as for 4U~1820--30 (Tarana et al. 2007). In any case, the spectra characteristics and models of \source\/ are different from the 4U~1820--30 ones. Also the large uncertainties on the value for the corona electron temperature do not allows to derive the presence of a cut-off. The lack of any firm detection in the radio band support this scenario though non-thermal process could not be ruled out. Indeed in any case, 4U~0614+091 also only an upper limit were derived in the radio band but later presence of jet was detected (Migliari et al. 2006, 2006b). 


We have also investigated the spectrum of 4U~1608--522 during a quiescent state with a luminosity higher than previous reported ones. The quiescent state has been modelled by a thermal NSA emission coming from the cooling NS and, for the first time, a power law with $\Gamma \simeq$ 3 was required. The thermal emission is due to the cooling NS (with R=10 km and M=1.4M$_\odot$) with an effective temperature of 0.12 keV, comparable to previous observed temperature of 0.17 keV \citep{rut99}. The nature of the power law tail is still unclear. This component was also observed as the dominant one above 2-3 keV in X-ray quiescent spectrum of transient NSs such as Aql~X-1, Cen~X-4 (Asai et al. 2006, Campana et al. 2000, Rutledge et al. 2000), KS~1731--260 \citep{rut2002}. It could be originated by a residual low level accretion onto the NS magnetosphere also in the quiescence state \citep{campanareview} or to an interaction of pulsar wind/surrounding gas such as for PSR~1259-63 \citep{tavani97}. This last case could be not applied to \source\/ that is not a pulsar like system. This power law component was never detected during previous quiescent spectrum of 4U~1608-522 suggesting that it appears only occasionally and suggesting a variability in the quiescent emission.

\acknowledgments
This research has made use of data obtained with {\it INTEGRAL\/} an international collaboration. The authors thanks M. Federici for the continuous effort to update the {\it INTEGRAL\/} archive and software in Rome and G. De Cesare and L. Natalucci for the scientific and data analysis support. The authors thanks also E. Jourdain for the information and discussion about the SPI data. Moreover we thanks Heinke C.O. for usefull discussion about the quiescent state spectrum observed. We acknoweledge the ASI financial support via grant ASI-INAF I/008/07 and I/088/06.


%
\begin{table*}[ht]
\begin{center}
\caption{The log of the data used for spectral fitting. The colors  are used in the light curves, Fig.\ \ref{plottwo}, and hardness--intensity plots, Fig.\ \ref{plottre}.
\label{tab_dataSET}}
\begin{tabular}{lccc}  
\tableline
\tableline
Data set & Time start (MJD)  & Instruments Exposure (ks) & Spectral state \\
\tableline
 {\it INTEGRAL\/} &        &  IBIS ; JEM-X  &  \\
\tableline
 S1 (light blue) & 53433.0 & 10.2 ; 10.7  &  hard         \\
 S2 (purple)     & 53435.2 & 56.5 ; 32.0  &  soft/intermediate\\
 S3 (red)        & 53438.0 & 64.5 ; 36.7  &  soft         \\
 S4 (green)      & 53462.7 & 74.6 ; 63.5  &  soft         \\
 S5 (blue)       & 53598.8 & 45.6 ; 53.7  &  hard         \\ 
\tableline
{\it BeppoSAX\/} &         & LECS ; MECS; PDS   &         \\
\tableline
 Spe1998         & 50872   & 30.4 ; 13.2 ; 14.1 &  soft   \\
 Spe2000         & 51757   & 47.8 ; 23.3 ; 22.7 &  quiescence \\
\end{tabular}
\vspace{-0.4cm}
\end{center}
\end{table*}

\begin{table*}[ht]
\begin{center} 
\caption{Spectral fitting results for the spectra of the banana states with {\it INTEGRAL\/} and {\it BeppoSAX\/}. \label{tabspe_soft}}
\begin{tabular}{llcccc}  
\tableline
\tableline
        &        &    \multicolumn{3}{c}{{\it INTEGRAL\/}}  & \multicolumn{1}{c}{{\it BeppoSAX\/}}  \\
 &    & S2 & S3  &  S4 & spe 1998   \\
\tableline
\texttt{CompTT+diskbb} &$kT_{0}$ (keV)             &1.1                   & 1.1                   & 1.0                   & 0.8$^{+0.1}_{-0.1}$ \\
             &$kT_{\rm e}$ (keV)         &5.8$^{+1.}_{-0.3}$ & 3.4$_{-0.2}^{+0.2}$ & 3.0$_{-0.1}^{+0.1}$ & 2.4$^{+0.1}_{-0.1}$ \\
             &$\tau$                     &2.3$^{+0.3}_{-0.3}$ & 3.6$_{-0.3}^{+0.3}$ & 4.3$_{-0.2}^{+0.2}$ & 6.5$^{+0.1}_{-0.1}$\\
             &norm$_{\rm CompTT}$        &0.1$^{+0.1}_{-0.1}$ & 0.6$^{+0.1}_{-0.1}$ & 1.2$_{-0.2}^{+0.1}$ & 1.1$^{+0.1}_{-0.1}$\\
             &$kT_{in}$ (keV)            &0.5$^{+0.2}_{-0.2}$ & 0.6$^{+0.3}_{-0.1}$ & 0.6$_{-0.3}^{+1.0}$ & 0.6$_{-0.1}^{+0.1}$\\
             &$\chi_r^2$($\nu$)          &0.99(36)               & 0.97(33)               & 0.99(33)               & 1.35(403) \\
\tableline
\texttt{reflect*gauss+CompTT+diskbb} &$kT_{0}$ (keV)& 1.2                  & 1.1                   & 1.0                   & 0.6$^{+0.1}_{-0.1}$ \\
                     &$kT_{\rm e}$ (keV) & 7.0$^{+1.9}_{-1.2}$ & 3.4$^{+0.1}_{-0.3}$ & 3.0$^{+0.1}_{-0.1}$ & 2.6$^{+0.1}_{-0.1}$\\
                     &$\tau$             & 1.7$^{+0.5}_{-0.5}$ & 3.4$^{+0.4}_{-0.3}$ & 4.1 $^{+0.2}_{-0.2}$& 5.7 $^{+0.1}_{-0.1}$\\
                     &norm$_{\rm CompTT}$& 0.1$^{+0.1}_{-0.1}$ & 0.5$^{+0.1}_{-0.1}$ & 1.1$^{+0.1}_{-0.1}$ & 1.2$^{+0.1}_{-0.1}$\\
                     &$kT_{in}$ (keV)    & 0.7$^{+0.1}_{-0.1}$ & 0.6$^{+0.2}_{-0.1}$ & 0.5$^{+0.8}_{-0.1}$ & 0.5$^{+0.1}_{-0.2}$\\
                     &$\Omega/2\pi$      & 0.4$^{+1.3}_{-0.6}$ & 0.2$^{+0.6}_{-0.4}$ & 0.3$^{+0.4}_{-0.2}$ & 0.5$^{+0.1}_{-0.2}$\\ 
                     &$\sigma_{Fe}$ (keV)  & -- & -- & -- & 1.2$^{+0.2}_{-0.5}$\\
                     &EW (eV)            & -- & -- & -- &208 \\
                  &$\chi_r^2$($\nu$)     & 0.94(35)               & 1.01(32)               & 0.99(32)               & 1.15(401) \\
\tableline
\end{tabular}
\end{center}
\end{table*}

\begin{table*}[ht]
\begin{center}
\caption{Spectral fitting results for the spectrum of the island state (S1+S5 data sets) with JEM-X and IBIS.
\label{tabspe_hard}}
\begin{tabular}{l|cc}  
\tableline
\tableline
Parameters & \texttt{CompTT+diskbb} & \texttt{CompPS+diskbb} \\
\tableline
 $kT_{0}$ (keV)         &1.2   & 1.1$\pm$0.5  \\
 $kT_{\rm e}$ (keV)     &62.6$^{+46.0}_{-28.8}$ & 38.5$\pm$9.7 \\
 $\tau$                 &0.4$^{+0.6}_{-0.3}$    & 2.6$\pm$0.8  \\
 $\Omega/2\pi$          & --                       & 0.1$\pm$0.2  \\
 norm                   &0.003$^{+0.002}_{-0.002}$ & 74.5$\pm$138.3 \\
\tableline
$kT_{in}$ (keV)          &0.4$^{+0.3}_{-0.1}$    & 0.5$\pm$0.6     \\
\tableline
$\chi_\nu^2$($\nu$)      &0.79(49)                  & 1.06(47)  \\
\tableline
\end{tabular}
\end{center}
\end{table*}

\end{document}